# Empirical model of campus air temperature and urban morphology parameters based on field measurement and machine learning in Singapore

Zhongqi Yu[a, #, *], Shisheng Chen[a, #], Nyuk Hien Wong[a, #], Marcel Ignatius[a], Jiyu Deng[a], Yueer He[a], Daniel Jun Chung Hii[a]

[a] Department of Building, National University of Singapore, 4 Architecture Drive, Singapore 117566.
* Corresponding author: Zhongqi Yu (Tel.: +65 85875096, Email: zhongqi5081@gmail.com);
Shisheng Chen (Email: bdgcs@nus.edu.sg); Nyuk Hien Wong (Email: bdgwnh@nus.edu.sg).
# Equal contribution

**Abstract**

The rising air temperature caused by Urban Heat Island (UHI) effect has become a problem for Singapore, it not only affects the thermal comfort of outdoor microclimate environment, but also increases the cooling energy consumption of buildings. As part of a multiscale and multi-physics urban microclimate model, weather stations were installed at 15 points within kent ridge campus of National University of Singapore (NUS) and continuously recorded the microclimate data from February 2019 to May 2019. A Geographical Information System (GIS) map and 3-dimensional (3D) model were constructed for extracting urban morphology parameters such as *BDG*, *PAVE*, *WALL* and *HBDG*. Through a site survey, *SVF* and *GnPR* were calculated.

By using multi-criteria linear regression and machine learning, this research investigated five regression models for prediction of outdoor air temperature (daily maximum ($T$max), minimum ($T$min), average ($T$avg), daytime average ($T$avg-day), nighttime average ($T$avg-night)) including linear regression (LR), k-nearest neighbours (KNN), support vector regression (SVR), decision tree (DT) and random forests (RF). The analysis of variables by best subsets regression showed greenery played crucial role in the mitigation of both daytime and night-time UHI. Pedestrian level wind flow was helpful in heat release in the daytime. High-rise buildings provided self-shadowing to reduce ambient air temperature but higher SVF was harmful to heat release in the night-time. For regression models, RF had the best predictive performance. The RF-based outdoor air temperature prediction models had RMSE range of 0.17°C to 0.55°C. Average RMSE of RF was reduced by 4% to 29% compared to linear regression. The learning curve indicated that the predictive power of LR could not be improved by additional data provision. In contrast, the downward trend in bias and variance suggested that RF can benefit from the training of big data. During the deployment of learning algorithms, RF continued to outperform other learning algorithms.

## 1 Introduction

Urbanized regions produce the majority of greenhouse gas emissions and are the places most vulnerable to human health impacts resulting from climate change (Bulkeley and Betsill, 2005; Revi et al., 2014). Cities are experiencing a higher rate of warming than proximate rural areas, the frequency, intensity and duration of heat waves to be increasing rapidly in many large cites (Vargo, 2016) in the context of global warming.

As a densely-populated city-state, Singapore has experienced rapid urbanization and fast economic growth since 1960's. Rising temperature has become one of the main environmental concerns. The annual average surface temperature in Singapore has increased from 26.6°C in 1972 to 27.7°C in 2014, it is predicted to rise by 1.4-4.6°C by 2099 (Meteorological Service Singapore (MSS), 2015).

This common problem, which named as Urban Heat Island (UHI) effect have attracted research attentions in many cities worldwide (Oke 1971; Padmanabhamurty 1990/91; Sani 1990/91; Swaid and Hoffman 1990; Eliasson 1996; Giridharan et al. 2007) as well as in Singapore. It is mainly caused by the replacement of natural surfaces by built structures, low wind speed condition, complex terrain, and anthropogenic heat (United State Environmental Protection Agency (EPA), 2008). Various models and tools have been proposed at different scales (Global/regional scale, urban scale and building scale).

At global/regional scale, some climatic models have been developed to simulate the turbulences and predict the climate change, such as the Regional Spectral Model (RSM) and Weather Research and Forecasting (WRF) model. At the building scale, the simulation of building energy consumption has been the main focus, e.g., EnergyPlus, Revit, IES-VE, and TRNSYS (Perez-Lombard, et al. 2008; Neto and Fiorelli, 2008). Between them, at the urban scale, both numerical and statistical approaches were adopted. Some computer programs have been developed to predict the microclimate temperature and air flow as numerical approach, such as ENVI-met, ANSYS Fluent, OpenFOAM, etc. (Wong, 2009; Gousseau et al., 2011; Yuan and Ng, 2012; Ng, et al., 2012). At the same time, based on field measurement, some empirical models were proposed as statistical approach (Svensson, 2002; Krüger, 2007; Tong, 2018). Closest related to this research, background studies (Wong et al. 2007; Jusuf et al. 2007; Jusuf and Wong, 2012) have been conducted and confirmed the existence of temperature patterns in relation to urban morphology conditions, by setting up weather stations at 2 sites in Singapore, the most significant parameters within 50m radius area were screened out and linear-regression equations for predicting air temperature were proposed.

The attention on these microclimate issues have helped to advance the development of urban and landscape planning and design strategies in Singapore. However, the linear regression used in most previous research is a parametric regression that is only useful when the relationship between the target variable and the input variables is known during the modelling process. In urban microclimate, the true form between urban morphology features and air temperature is unclear. Therefore, non-parametric regression such as machine learning is well suited for outdoor temperature prediction because it does not assume a predictive model of any shape but rather relies on the data information provided. This research evaluated four machine learning algorithms including k-nearest neighbours (KNN), support vector regression (SVR), decision tree (DT) and random forests (RF) for prediction of outdoor air temperature. The assessment was based on cross-validation rather than train and test split providing much objective and repeatable predictive performance. R version 3.5.2 was used for variable selection and best subset regression (R Core Team, 2018). The training and testing of machine learning algorithms were implemented on Python version 3.7.3 and open source machine learning library Scikit-Learn version 0.2 (Python Software Foundation, 2019; Pedregosa et al., 2011).

## 2 Objectives and scope of work

As part of virtual campus project, this research is aim to adopt machine learning on microclimate and reveal the relationship between campus air temperature and urban morphology indicators in Singapore with better accuracy. By conducting the following works, key variables in the study of UHI effect at NUS campus were identified, and the most appropriate regression model for outdoor air temperature prediction were determined. It could be applied to optimizing existing urban built environment, assessing new building and proposing design codes for future urban planning:

- Field measurement at 15 points around NUS kent ridge campus to collect microclimate data for 3 months from February 2019 to April 2019, 14 points for developing model, 1 point for collecting reference temperature data;
- Develop a multiscale and Multi-physics urban microclimate model for NUS campus including GIS map and 3D model, and extracted urban morphology parameters;

- Develop empirical models with multi-criteria linear regression and machine learning to correlate the campus 2.4m height air temperature at each point with their morphology parameters and reference weather data;
- Statistical comparison multi-criteria linear regression and machine learning regression models for exploring higher accuracy potential of microclimate air temperature prediction.

## 3 Methodology
### 3.1 Field measurement

15 weather stations were set up across the NUS Kent Ridge Campus to collect the microclimate data. Apart from an evenly distribution, these weather station locations were selected for capturing typical urban features, such as urban canyon, dense tree area, open field area and campus highest point. These locations also include overheated points in NUS indicated by previous studies.

As shown in **Figure 1**, each weather station with 4 sensors and data loggers powered by solar panels were installed at 2.4m height to record 6 parameters data (air temperature, relative humidity (RH), global solar radiation, wind speed, wind direction (2D) and wind gust) at 1-minute interval from Feb 1, 2019 to April 30, 2019. Specifications for these sensors are listed in **Table 1**. Instead of traditional on-site data collection mode, the data collected in this project were transferred to cloud storage via 3G wireless connection.

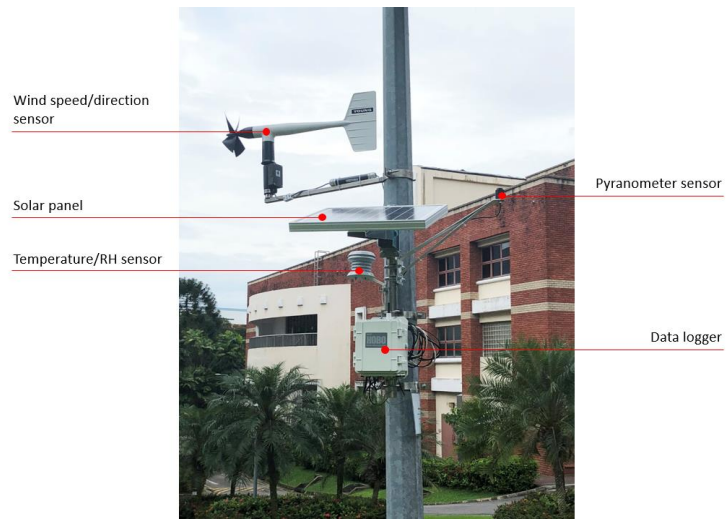

**Figure 1:** Installed weather station example

Table 1: Specifications of field measurement sensors

| Parameter | Temperature | RH | Solar Radiation | Wind Speed | Wind Direction |
|---|---|---|---|---|---|
| Range | -40°C-75°C | 0-100% | 0-1280W/㎡ | 0-50m/s | 0-360° |
| Accuracy | ±0.21°C | ±2.5% | ±10W/㎡ | 0.2m/s | 1.4° |
| Resolution | 0.01°C | 0.1% | 1.25W/㎡ | 0.2m/s | 1.4° |

At the same time, a set of reference data was continuously collected from PGP block 2 rooftop. Later, these data were used in 2 ways: as indicator for screening campus weather data samples and as weather parameters for developing air temperature empirical models.

### 3.2 Microclimate data selection

This research is focus on empirical model between campus air temperature and urban morphology parameters. Thus, only days with fairly clear, calm and hot weather conditions during our measurement period were reasonably selected for data analysis and model development (Tong, 2017).

Based on previous long-term observation and study, the fairly clear and sunny days with average temperature higher than 22°C were selected, rainy and cloudy days were excluded from data base. The following criteria were proposed to select ideal typical sample days.

- Daily maximum global solar radiation larger than 800W/㎡;
- Daily average temperature higher than 22°C;
- Daily average wind speed lower than 3m/s and no rain;
- Bell-shape hourly temperature profile and global solar radiation profile.

**3.3 GIS map and 3D model**

In this research, the campus air temperature was assumed influenced by urban morphology features within 50m radius area (Jusuf, 2009). Based on the project cooperation platform, a GIS map was developed in ArcGIS pro and a 3D model was constructed in STL format to represent the campus urban morphology especially near our field measurement points, as illustrated in **Figure 2**. The GIS map contains geographic information of building footprint, road, and pavement. The 3D model finely represented buildings and terrain in kent ridge campus in detail.

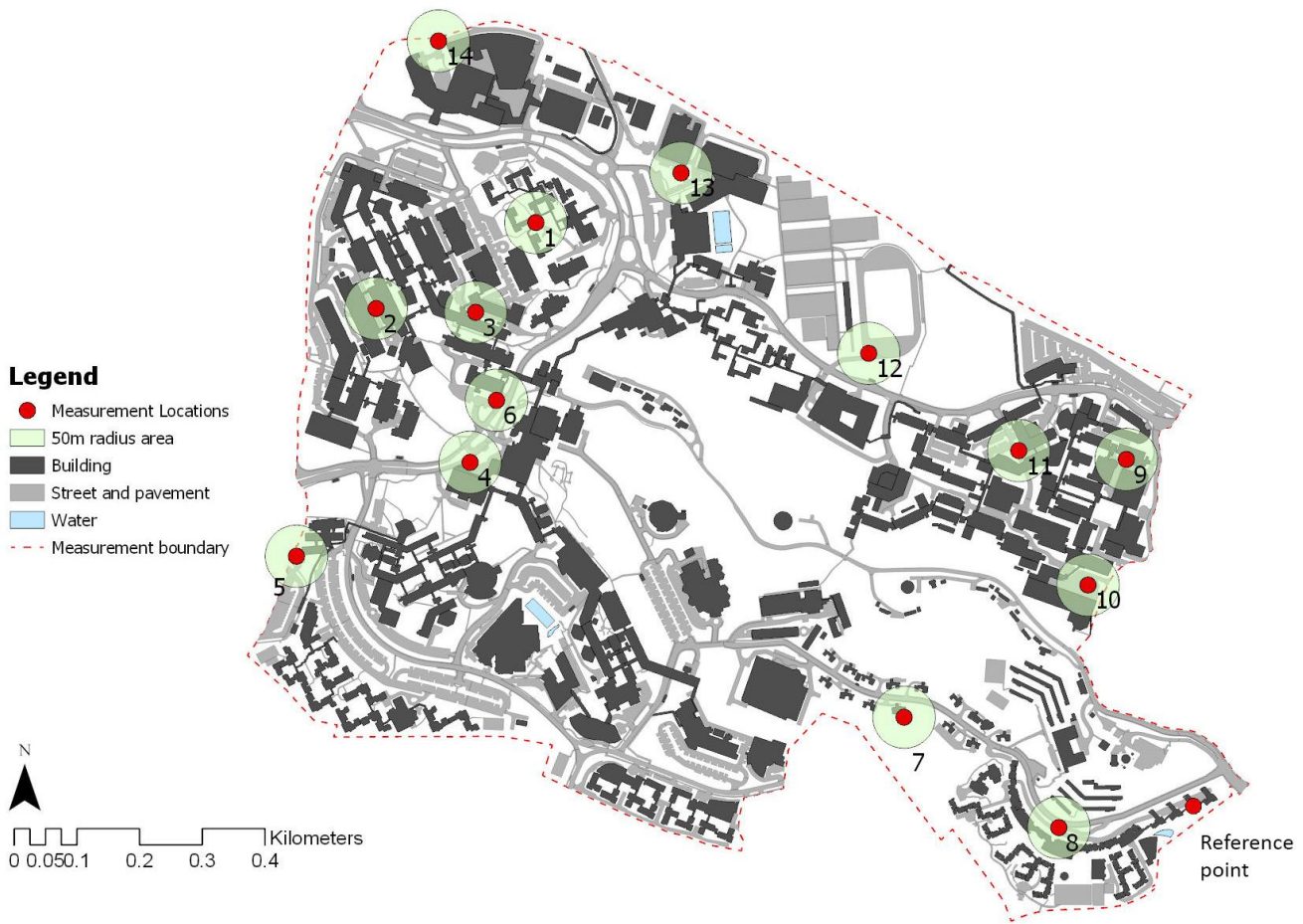

**Figure 2:** GIS map of NUS Kent Ridge campus

**3.4 Urban morphology analysis**

5 urban morphology parameters around 14 meteorological stations were extracted from the GIS map and 3D model as below:

- *BDG*: percentage of building area within 50m radius;
- *PAVE*: percentage of pavement area within 50m radius.
- *H*: average building heights within 50m radius;
- *WALL*: total exterior wall surface area, in m²;
- *HBDG*: ratio of average building heights over the percentage of building footprint, in m.

Among them, *BDG*, *PAVE* and *H* were extracted from GIS map, and *WALL* was measured and calculated from 3D model. In particular, *HBDG* represents the overall thermal mass was calculated from both GIS map and 3D model.

Besides, in April of 2019, a site survey was conducted to identify greenery condition and *SVF* within 50m vicinity around 14 weather stations. As shown in **Figure 3**, the SVF of each measurement point was measured by means of Nikon SLR camera with fish eye lens. The trees, shrub and turf area as well as species and location of trees were surveyed and mapped for calculating *GnPR*.

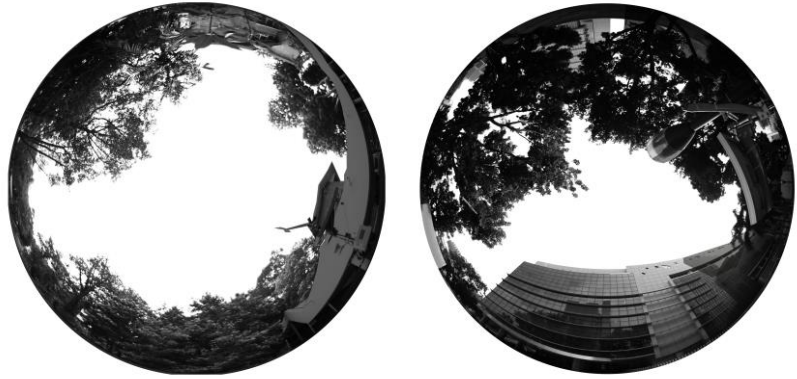

**Figure 3:** Fish-eye photos of point 1 and 4

The *SVF* was calculated from fish-eye photos. The *GnPR* proposed by Ong is the area-weighted average leaf area index of a site. In practice, in the calculation of *GnPR*, the leaf area index (LAI) of each species would be multiplied by the canopy area or planted area and the total for all species in the site would be divided by the site area. In this work, the *GnPR* can therefore be calculated as (Tan, 2017):

$$GnPR = \frac{total\ leaf\ area}{site\ area}$$
$$= \frac{\sum(LAI_1 \times Canopy\ Area_1 + LAI_2 \times Canopy\ Area_2 \cdots \cdots + LAI_n \times Canopy\ Area_n)}{(Site\ Area)} \quad (1)$$

Where the canopy area was calculated respectively for various species of trees, shrub and turf. Crown of tree was simplified as a circle. Number of each tree species were counted and their crown radius were measured. For trees, LAI from 2.5 to 4.0 were assigned based on dense, intermediate or open canopy. Palms were given special value. For shrub, monocot and dicot shrub were counted respectively.

### 3.5 Variable selection and model validation

Identifying the most relevant variables is crucial in this research, providing useful insights for urban heat island on the NUS campus and design guidelines for future development. Prior to comparing linear regression and machine learning algorithms, best subset regression (BSR) implemented in R was used to find the optimal variables from all possible subsets. Compared with BSR, variable selection based on simple F test or full model t test is equivalent to discarding or adding servo variables at the same time, which may miss the optimal model. Given today's computing power, a safe approach is to implement BSR during variable selection. Bayesian information criterion (BIC) was selected as evaluation metric for variable selection because it is function of error variance and parameter number shown in Equation 2. Finding the best model is a trade-off between bias and variance, complex model provides low bias but high variance results. A best model is a smallest model fits the data. Hence, variable selection based on BIC will meet the requirement of searching best model for least error variance and simplest form. This research reported BIC value as the difference between model with best variables and model with intercept only to indicate the improvement made by best subsets model. In addition, adjusted R-squared ($R_a^2$) was reported to show the predictive power of each air temperature prediction model.

$$\text{BIC} = n \cdot \log\left(\frac{\text{SSE}}{n}\right) + p \cdot \log(n) \qquad (2)$$

where
- n: number of observations;
- p: number of parameters;
- SSE: sum of squared error is defined as sum of difference between true $Y_i$ and fitted $\hat{Y}_i$, $\text{SSE} = \sum_{i=1}^{n}(Y_i - \hat{Y}_i)^2$.

In terms of model validation, 10 repeats 5 folds cross-validation was applied during the modelling to provide unbiased error estimation. All labelled data were divided into 5 folds and each fold had opportunity to be used as testing data to validate the predictive model. The division of 5 folds was processed repeatedly by 10 different approaches. Root mean square error (RMSE) shown in Equation 3 was adopted as error metric of model because it provides same unit with target variable and penalizes heavily on large error than mean absolute error (MAE). Cross validation was applied to both linear regression and machine learning algorithm to make fair comparisons of regression models during the learning process.

$$\text{RMSE} = \sqrt{\frac{1}{n}\sum_{i=1}^{n}(Y_i - \hat{Y}_i)^2} \qquad (3)$$

Where
- n: number of observations;
- $Y_i$: True target variable;
- $\hat{Y}_i$: Estimated target variable.

**3.6 Linear regression and machine learning**

In python, linear regression (LR) was implemented by Scikit-learn algorithm "LinearRegression". As for machine learning, outdoor air temperature prediction was a supervised regression problem. Therefore, this research selected four machine learning algorithms for evaluation including k-nearest neighbours (KNN), support vector regression (SVR), decision tree (DT) and random forests (RF). The settings of regression models are illustrated in **Table 2**.

**Table 2:** Settings of regression model and algorithm

| Regression model | Scikit-learn algorithm | Settings |
| --- | --- | --- |
| Linear regression (LR) | LinearRegression | N/A |
| K-Nearest Neighbors (KNN) | KNeighborsRegressor | Number of neighbors=5; Weight function=uniform; Distance function=Euclidean distance |
| Support Vector Regression (SVR) | SVR | Kernel=rbf; C=1; Epsilon=0.1 |
| Decision Tree (DT) | DecisionTreeRegressor | Min samples split=2; Min sample leaf=1 |
| Random Forests (RF) | RandomForestRegressor | Number of estimators=100; Min samples split=2; Min sample leaf=1 |

KNN estimates new prediction by local interpolation of nearest neighbours based on feature similarity from training data (Harrington, 2012). The feature similarity is calculated by the distance function between new point and each of training data. The prediction of new data is the average of the k neighbours.

SVR is an epsilon support vector regression based on LIBSVM with two free parameters C and epsilon for tuning (Scikit-learn developers, 2018a). The idea behind SVR is to map the input variables to high dimensional feature space by kernel function and use linear hyperplane for separation (Cortes and Vapnik, 1995). SVR is stopped by epsilon loss function when the difference between prediction and actual value is less than or equal to epsilon.

Scikit-learn adopts classification and regression tress (CART) algorithm inside the decision tree regressor (Scikit-learn developers, 2018b). CART predicts target variable based on decision rule to best split the observations (Breiman, 2017). The rule chooses input attribute with greatest error reduction in target variable at each node. The process is recursive until meeting the stopping criterion. The tree is then pruned back to overcome overfitting problem. Random forest (RF) is a combination of tree predictors which splits node based on best split of random subsets of the features thus reducing the variance of tree model and increasing overall predictive power of model (Breiman, 2001).

## 4 Result and discussion
### 4.1 Weather data selected

According to the stipulated criteria described in **Section 3.2**, as illustrated in **Table 3**, 44 typical days from February 2019 to May 2019 were selected for analysis, among them, 39 days from February to April were used for model development and cross validation, and 5 days in May were used for model deployment test. Due to short-term error or maintenance, some days with no readings or extremely unstable readings were also excluded from the data set. The selected days from February to May were randomly divided into groups for model development and validation respectively.

**Table 3:** Typical sample days selected

| Month | February 2019 | March 2019 | April 2019 | May 2019 |
|---|---|---|---|---|
| **Typical Days selected** | 2, 3, 5, 6, 7, 9, 16, 21, 22, 23, 24, 25, 26, 27, 28 | 1, 2, 3, 6, 10, 14, 15, 17, 18, 19, 22, 26, 27, 29 | 6, 7, 8, 12, 13, 15, 17, 18, 21, 29 | 21, 24, 26, 29, 31 |

### 4.2 Urban morphology parameters

Urban morphology parameters *BDG*, *PAVE* were calculated for 50m radius area of each measurement points in ArcGIS pro, *WALL* was measured in 3D model. For calculating *HBDG*, height of each building was measured in 3D model and building footprint area percentage was measured in ArcGIS pro. *SVF* at each point was estimated by analyzing fisheye photo in PC program RayMan. For calculation *GnPR*, various values were assigned to greenery depend on its species. 4.0, 3.0 and 2.5 were respectively assigned to trees with dense canopy, intermediate canopy and open canopy. For palms, 2.5 and 4.0 were assigned to single-stemmed and multi-stemmed palm. 3.5 and 4.5 were given to monocot and dicot shrub. Finally, 2.0 was assigned to turf as LAI value (Tan, 2017). The results of urban morphology parameters of each point were presented in Appendix A.

## 4.3 Variable selection and linear regression

Table 4: Variable selection of outdoor air temperature models

| Variable | $T_{max}$ | $T_{avg}$ | $T_{min}$ | $T_{day\text{-}avg}$ | $T_{night\text{-}avg}$ |
|---|---|---|---|---|---|
| Intercept | 10.87 | 1.3 | 1.69 | 5.2 | 1.14 |
| SVF | 0.73 | 0.78 | -0.43 | 1.41 | NA |
| HBDG | NA | NA | NA | NA | NA |
| WALL | NA | 2.64E-05 | 1.79E-05 | 3.14E-05 | 1.20E-05 |
| GnPR | -0.26 | -0.11 | -0.29 | NA | -0.3 |
| BDG | NA | NA | NA | NA | NA |
| PAVE | NA | 1.07 | 0.5 | 1.66 | NA |
| $T_{ref}$ | 0.67 | 0.93 | 0.95 | 0.78 | 0.98 |
| $Solar_{avg}$ | 9.57E-03 | 1.28E-03 | NA | 3.66E-03 | NA |
| $Wind_{avg}$ | -1.48 | -0.23 | NA | -0.69 | NA |
| BIC | -1000 | -1700 | -1900 | -1500 | -2100 |

Based on the multicollinearity test and BSR, Table 4 summarizes the optimal variable selection for the five temperature models. Although their effects are different, certain urban morphology variables (e.g. SVF, WALL, GnPR, PAVE) are common in temperature models.

SVF is positively correlated with daytime temperatures (e.g. Tmax, Tavg, Tday-avg), and negatively correlated with nighttime Tmin. This means that low SVF from high-rise buildings will reduce ambient air temperature by blocking sunlight, but low SVF will cause obstacles to nighttime long-wave radiation release. In general, a low SVF caused by trees generates more optimistic thermal environments (Zhang et al., 2019). WALL is considered to be an active contributor to the temperature indices (e.g. Tavg, Tmin, Tday-avg, Tnight-avg) because of the high emissivity of building materials. Similarly, PAVE is positively correlated with Tavg, Tmin and Tday-avg due to its high emissivity. In contrast, the increase in GnPR leads to a reduction in Tmax, Tavg, Tmin and Tnight-avg, as greenery provides shading and evapotranspiration during the day and it does not reflect long-wave radiation at night. Windavg is negatively correlated with Tmax, Tavg and Tday-avg, indicating that strong winds play a vital role in easing the temperature rise during the day at the NUS campus.

As illustrated in **Figure 4,** although the accuracy of linear regression (LR) was improved by more training data, green line of test error showed LR had some noteworthy fluctuation during the learning process. The learning curves of five air temperature models were converged or closed to convergence at maximum training size suggesting the predictive model could not be further improved by adding more training sample. In addition, linear regression does not offer hyperparameters for algorithm improvement.

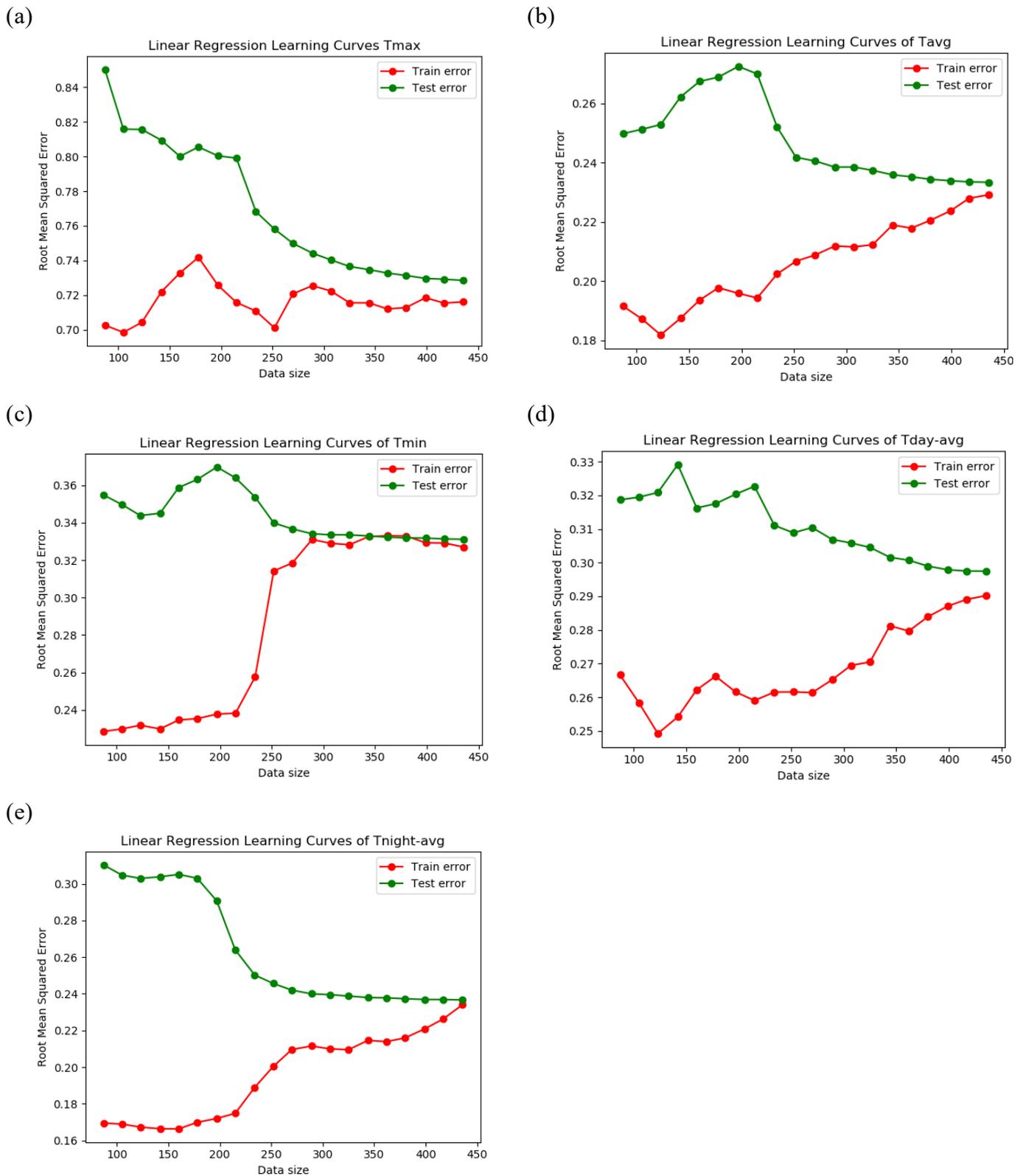

**Figure 4:** Learning curves of linear regression: (a) Tmax; (b) Tavg; (c) Tmin; (d) Tday-avg; (e) Tnight-avg

### 4.4 Comparison of machine learning algorithms

**Figure 5** illustrates the learning process of five regression algorithms for outdoor air temperature prediction. Overall, blue line of linear regression was superior than machine learning algorithms at small training size but surpassed by purple line of random forest at large training size. Red line of k-nearest neighbours (KNN) showed worst performance among the regression algorithms for outdoor air temperature prediction followed by orange line of support vector regression (SVR) and decision tree (DT).

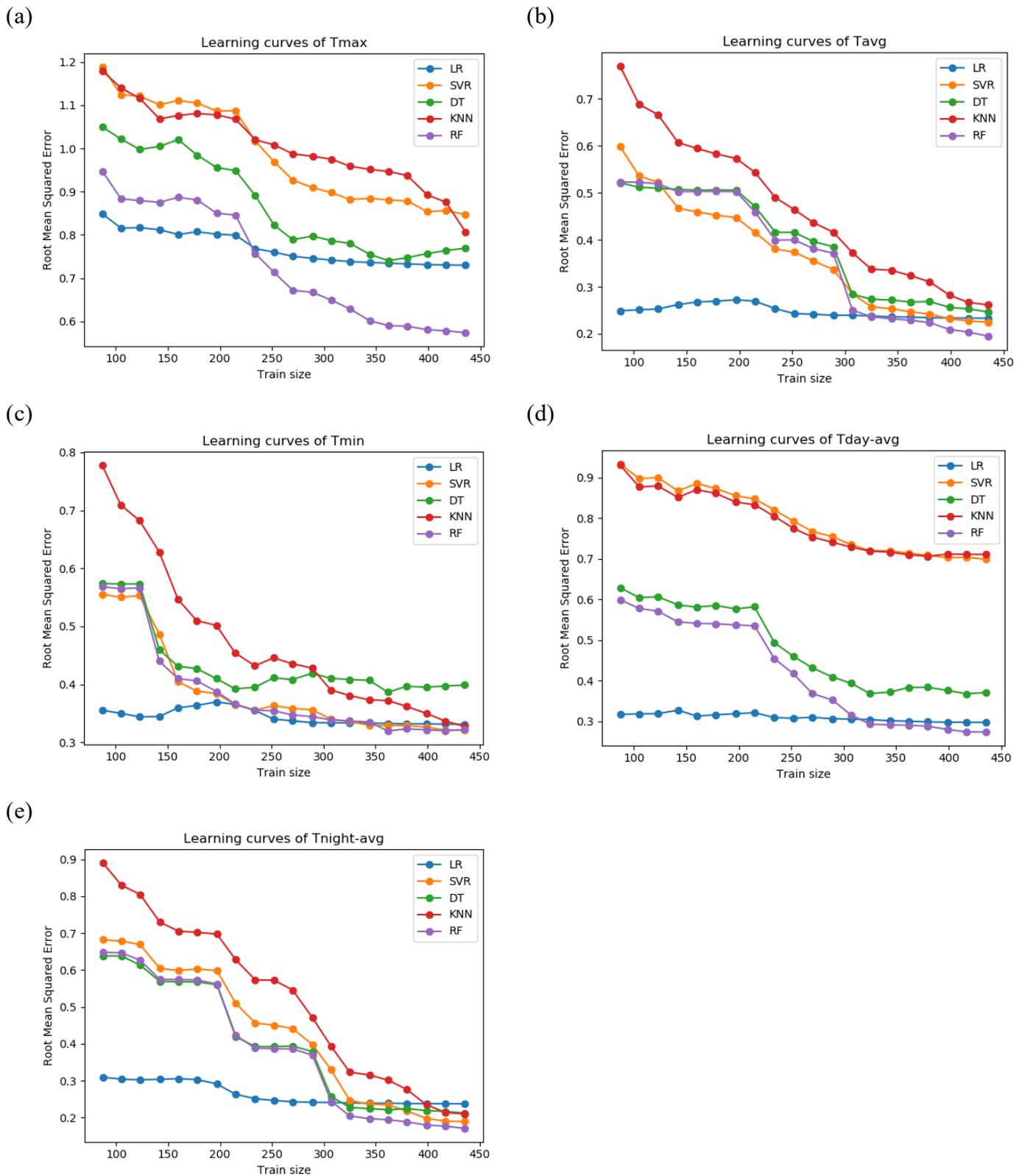

**Figure 5:** Comparison of test error among five regression algorithms for outdoor air temperature prediction:
(a) Tmax; (b) Tavg; (c) Tmin; (d) Tday-avg; (e) Tnight-avg

The performance of five regression algorithms was much diverged in $T$max and $T$day-avg prediction especially for KNN and SVR. In conclusion, since KNN uses full training data to predict new points, it is expensive to deploy an outdoor air temperature predictive model using KNN especially when KNN does not generate satisfactory results. SVR might compete with RF in part of air temperature prediction tasks but it is not as stable as RF. RF is superior than DT because it is made of multiple tree predictors and enhanced to overcome overfitting problem. Given the above analysis, RF is considered as optimal algorithm for outdoor air temperature prediction.

In comparison, random forest has ability to be trained and improved at large data sample. **Figure 6** states the bias of random forest predictive model was progressively reduced when model was trained with large sample. Meanwhile, the error variance of model has been narrowed down during the training. Because of complexity of random forest, simple linear regression model generated much lower error variance than random forest. However, the data sample in this research was relatively small with 546 samples. Linear regression will face challenges in dealing with big data sample because model bias cannot be reduced any further as highlighted before. The learning curves of random forest model were far from convergence at maximum training size thus model accuracy can be improved by adding more training sample. Moreover, there is an opportunity for hyperparameter turning in random forest.

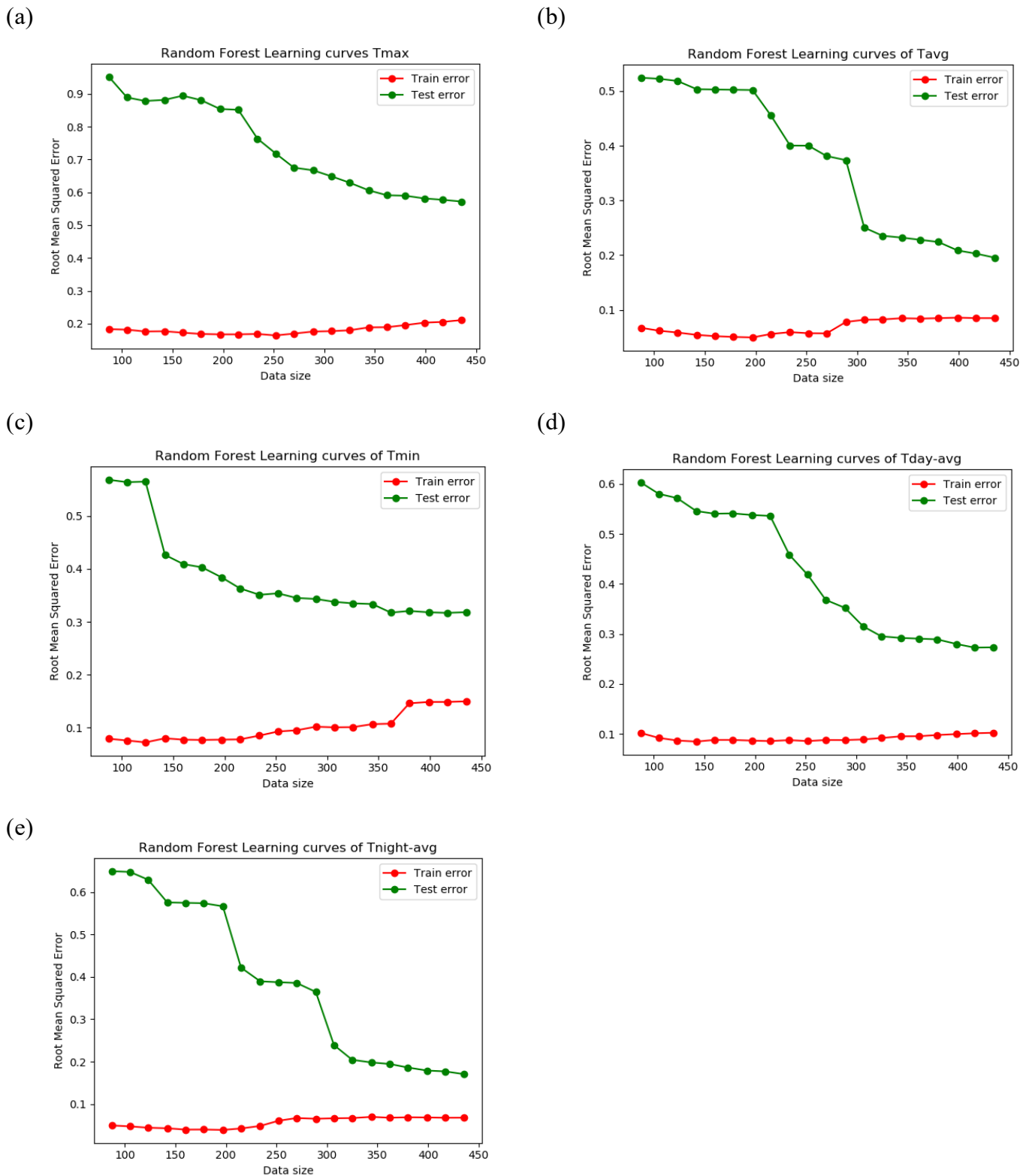

**Figure 6:** Learning curves of random forest: (a) Tmax; (b) Tavg; (c) Tmin; (d) Tday-avg; (e) Tnight-avg

**Table 4:** Comparison of root mean squared error (RMSE) between random forest and linear regression for outdoor air temperature prediction

| Target | Algorithm | RMSE | | | | (LR-RF)/LR | | | |
|---|---|---|---|---|---|---|---|---|---|
| | | Mean | Std | Max | Min | Mean | Std | Max | Min |
| Tmax | RF | 0.55 | 0.03 | 0.62 | 0.47 | 10% | 5% | 18% | -5% |
| | LR | 0.61 | 0.03 | 0.67 | 0.54 | | | | |
| Tavg | RF | 0.19 | 0.02 | 0.23 | 0.16 | 17% | 6% | 28% | -2% |
| | LR | 0.23 | 0.01 | 0.27 | 0.19 | | | | |
| Tmin | RF | 0.32 | 0.03 | 0.38 | 0.26 | 4% | 7% | 17% | -13% |
| | LR | 0.33 | 0.03 | 0.40 | 0.28 | | | | |
| Tday-avg | RF | 0.28 | 0.03 | 0.34 | 0.21 | 8% | 6% | 20% | -4% |
| | LR | 0.30 | 0.02 | 0.34 | 0.25 | | | | |
| Tnight-avg | RF | 0.17 | 0.02 | 0.23 | 0.13 | 29% | 7% | 43% | 6% |
| | LR | 0.24 | 0.02 | 0.28 | 0.20 | | | | |

**Figure 7** illustrates the simulation results of air temperature prediction based on 50 times validations. The red dot line and green dot line represent RMSE of linear regression (LR) and random forests (RF), respectively. The purple bar stands for the change between LR and RF at each validation set. The numerical results are summarized in **Table 4**. Overall, average performance of RF showed dramatic improvement than LR in prediction $T$max, $T$avg and $T$night-avg prediction with average RMSE improved by 10%, 17% and 29%, respectively. Prediction of $T$min and Tday-avg had slight increase of 4% and 8% by RF over LR, respectively. During cross validation, RF consistently outperformed LR in $T$night-avg prediction. In contrast, $T$min tended to have more variation during the cross validation.

The RF-based outdoor air temperature prediction models had RMSE range of 0.17°C to 0.55°C. The standard deviation of RMSE of RF was slightly higher than linear regression during the cross-validation. This is due to the complexity of random forest that complex regression model tends to have higher variance predictive results. As long as the bias of model is continuously decreased during the learning process, high variance is acceptable. Among the five target variables, the $T$max regression model has the highest RMSE which is approximately twice the other target variables. The low predictive power of $T$max model was consistent with previous research conducted by Steve for estate level air temperature prediction in Singapore (Jusuf and Wong, 2009). This means that in addition to the urban morphology and weather data, there are factors (e.g. anthropogenic heat) that strongly influence $T$max.

(a)  (b)

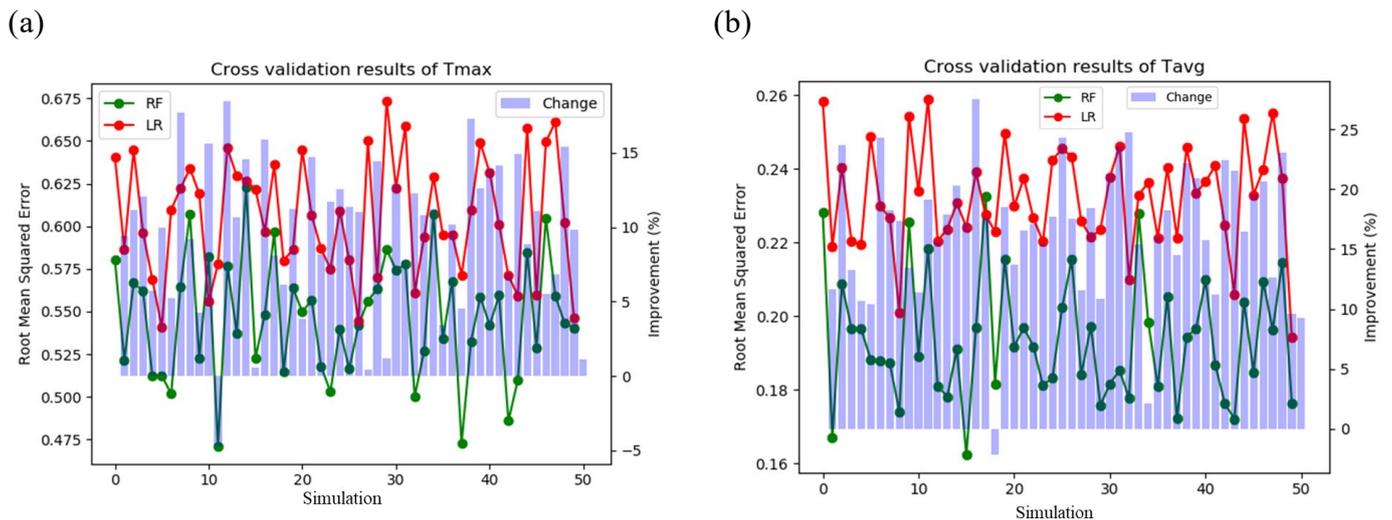

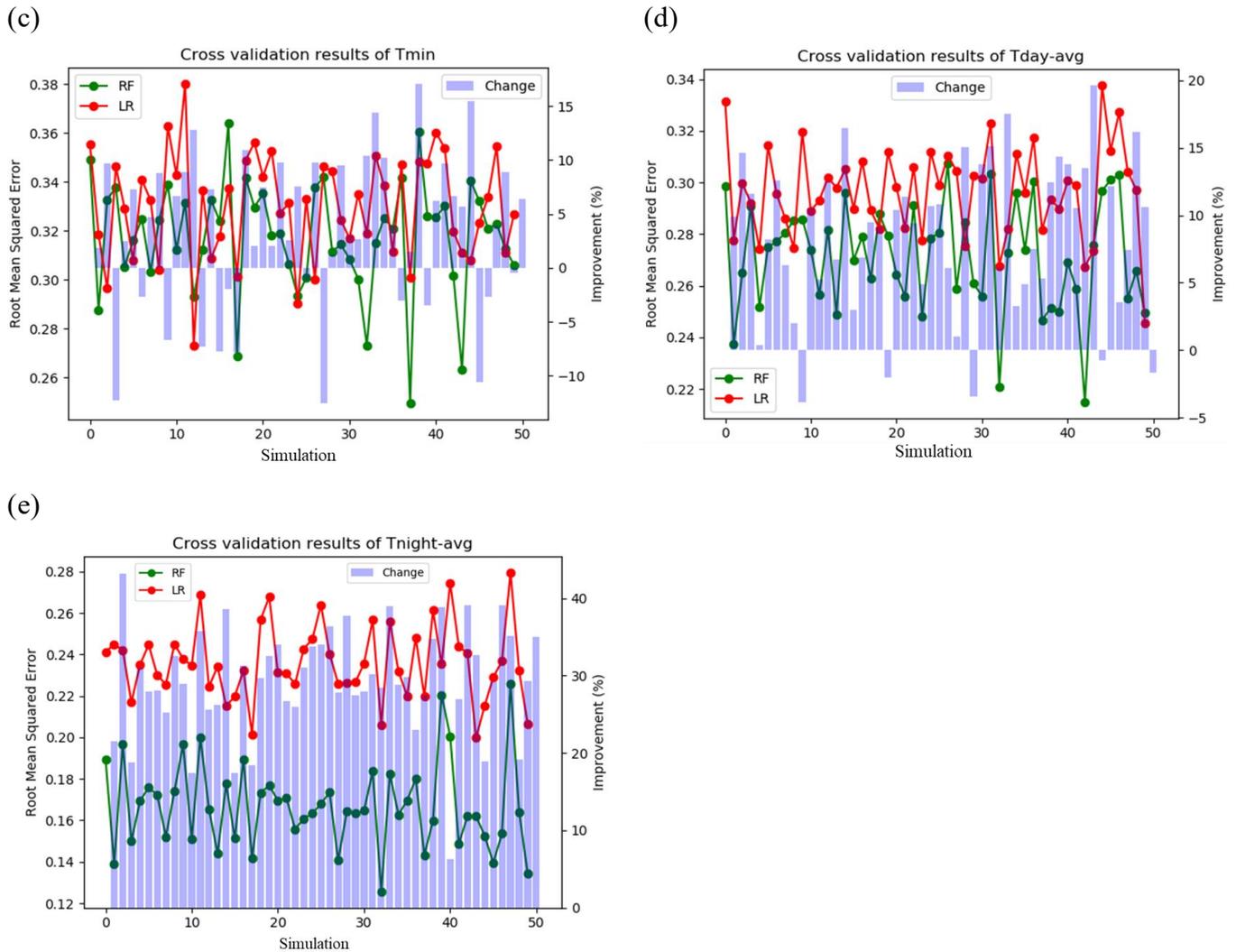

**Figure 7:** Cross validation results of outdoor air temperature prediction: (a) Tmax; (b) Tavg; (c) Tmin; (d) Tday-avg; (e) Tnight-avg

### 4.5 Deployment of Machine Learning Models

Weather data in the month of May has been selected to test the performance of the post-deployment machine learning models. The results shown in **Table 6** indicated the performance of learning algorithms for the new data. Overall, RF continuously outperformed other learning algorithms including LR, SVR, KNN and DT in the prediction of outdoor air temperature. The RF-based outdoor air temperature prediction model had a higher RMSE range of 0.25 °C to 0.97 °C than the cross-validation prediction.

When predicting $T$avg, $T$day-avg and $T$night-avg, RF was superior to LR despite their lower performance than expected but within the forecast of cross-validation as shown in **Table 5**. Compared with linear regression, the RMSE of $T$avg, $T$day-avg and $T$night-avg in RF during the deployment decreased by 15%, 6% and 15%, respectively. This demonstrated that the variables selected from best subsets regression for the three models were sufficient to form a high prediction model. As for $T$min, RF performance was 12% lower than LR, which was better than expected. RF did not perform better than LR in predicting $T$max, which again proved that $T$max may be ineffective in predicting new data because of the low predictive power and other factors that cause temperature variations in $T$max.

Table 5: Comparison of root mean squared error (RMSE) among learning algorithms after the deployment

| Target | LR | SVR | KNN | DT | RF | (LR-RF)/LR |
|---|---|---|---|---|---|---|
| Tmax | 0.94 | 1.25 | 1.16 | 1.10 | 0.97 | -2% |
| Tavg | 0.36 | 0.33 | 0.34 | 0.31 | 0.30 | 15% |
| Tmin | 0.50 | 0.55 | 0.52 | 0.46 | 0.44 | 12% |
| Tday-avg | 0.47 | 0.86 | 0.76 | 0.48 | 0.45 | 6% |
| Tnight-avg | 0.29 | 0.25 | 0.24 | 0.28 | 0.25 | 15% |

## 5 Conclusion

This work proposed and compared empirical models developed from linear regression and machine learning between urban morphology parameters and campus air temperature, answered two fundamental questions that predict outdoor air temperature on the campus of the National University of Singapore, including which parameters should be considered as key variables and which regression model should be the most appropriate.

The analysis of variables showed greenery played crucial role in the mitigation of both daytime and night-time UHI. Pedestrian level wind flow was helpful in heat release in the daytime. High-rise buildings provided self-shadowing to reduce ambient air temperature but higher *SVF* was harmful to heat release in the night-time.

As for regression model, random forests (RF) has best performance in terms of data fit. Linear regression (LR) outperformed k-nearest neighbour (KNN), support vector regression (SVR) and Decision tree (DT). The outdoor air temperature prediction model based on RF had RMSE ranging from 0.17°C to 0.55°C. *T*max model had lowest predictive power compared to other target variables suggesting missing of factors in the model. By using RF as a regression model, *T*max, *T*avg and *T*night-avg had relatively large RMSE reductions of 10%, 17% and 29%, respectively. There was medium decrease of RMSE of 8% in *T*day-avg.

The converged learning curves of linear regression showed the model could not be enhanced further by additional data training. Conversely, RF model had declining trend of bias and variance when training sample increased. During the deployment, RF continued to outperform other learning algorithms, including LR, SVR, KNN, and DT. RF is better than LR when predicting *T*avg, *T*day-avg and *T*night-avg, although their performance was lower than expected but within the cross-validation prediction range.

This research has some limitations regarding the data samples and measurement. Due to limited measurement duration and some raining days from February to April, only 546 samples were used in analysis, as the measurement continues, the models can be further improved. The low prediction power for *T*max also indicates that there are other factors influence the result, which is still subject to further experiment.

**Acknowledgement**

This work was supported by NUS internal project (WBS R-296-000-186-133). Besides, the authors sincerely appreciate kindly advices from Tong Shanshan for methodology and dedicated work from all team member during the weather station installation as well as monitoring.

**Appendix A**

Urban morphology predictors within 50m radius area of 14 measurement points

| No. | BDG | PAVE | H (m) | WALL (㎡) | HBDG (m) | SVF | GnPR |
|---|---|---|---|---|---|---|---|
| 1 | 0.258 | 0.099 | 10.250 | 12004.200 | 0.397 | 0.422 | 2.846 |
| 2 | 0.648 | 0.132 | 13.980 | 25690.300 | 0.216 | 0.365 | 0.459 |
| 3 | 0.377 | 0.205 | 25.050 | 21692.400 | 0.664 | 0.164 | 1.453 |
| 4 | 0.326 | 0.327 | 25.910 | 14387.500 | 0.795 | 0.376 | 1.995 |
| 5 | 0.130 | 0.373 | 19.100 | 5596.700 | 1.469 | 0.527 | 2.029 |
| 6 | 0.190 | 0.374 | 20.480 | 9115.000 | 1.078 | 0.480 | 1.915 |
| 7 | 0.032 | 0.137 | 6.730 | 2259.060 | 2.103 | 0.415 | 3.640 |
| 8 | 0.262 | 0.292 | 32.740 | 25072.100 | 1.250 | 0.471 | 1.630 |
| 9 | 0.584 | 0.211 | 37.980 | 28503.270 | 0.650 | 0.340 | 0.524 |
| 10 | 0.400 | 0.248 | 37.020 | 9682.340 | 0.926 | 0.263 | 1.535 |
| 11 | 0.498 | 0.195 | 22.420 | 21919.000 | 0.450 | 0.183 | 1.042 |
| 12 | 0.009 | 0.387 | 3.600 | 68.200 | 4.000 | 0.748 | 1.593 |
| 13 | 0.301 | 0.378 | 17.100 | 9343.800 | 0.568 | 0.605 | 1.370 |
| 14 | 0.137 | 0.332 | 26.960 | 9218.320 | 1.968 | 0.346 | 2.068 |

**Nomenclature**

GIS: Geographical information system
UHI: Urban heat island
NUS: National University of Singapore
*BDG*: Percentage of building area within 50m radius area
*PAVE*: Percentage of pavement area within 50m radius area
*WALL*: Total exterior wall surface area within 50m radius area ($m^2$)
*HBDG*: Ratio of average building heights over the percentage of building footprint (m)
*SVF*: Sky view factor
*GnPR*: Green plot ratio
*Tmax*: Daily maximum temperature (℃)
*Tmin*: Daily minimum temperature (℃)
*Tavg*: Daily average temperature (℃)
*Tavg-day*: Daytime average temperature (℃)
*Tavg-night*: Night-time average temperature (℃)
*Windavg*: Daily average wind speed (m/s)
*Solaravg*: Daily average global solar radiation (W/$m^2$)
LR: Linear regression
KNN: K-nearest neighbours
SVR: Support vector regression
DT: Decision tree
RF: Random forests
SLR: Single lens reflex camera
LAI: Leaf area index
BSR: Best subset regression
RMSE: Root mean square error
BIC: Bayesian information criterion